\documentclass[12pt]{iopart}
\usepackage[T1]{fontenc}
\usepackage{graphicx}
\usepackage{dcolumn}
\usepackage{bm}
\usepackage{amssymb}
\usepackage[dvipsnames]{xcolor}
\usepackage[bottom=1in]{geometry}
\expandafter\let\csname equation*\endcsname\relax
\expandafter\let\csname endequation*\endcsname\relax
\usepackage{amsmath}
\usepackage{appendix}
\begin{document}

\title[]{Post-transcriptional Regulation of Stochastic Gene Expression Conditioned on Large Deviations}

\author{Kenny Wong$^{1,\dagger}$, Thomas Mourier$^2$, \& Cameron Hopkinson$^3$}
\address{$^1$Department of Physics, The Ohio State University, Columbus, Ohio 43210 \\}
\address{$^2$Department of Mechanical and Industrial Engineering, Northeastern University, Boston, MA 02115 \\}
\address{$^3$Department of Biology, Washington University, St. Louis, MO 63130 \\}
\address{$^\dagger$\textbf{Email:} \texttt{wong.1216@osu.edu}}

\begin{abstract}
    Gene expression is a stochastic process that gives rise to large fluctuations in protein levels leading to phenotypic heterogeneity in clonal cell populations; post-transcriptional regulation plays a crucial role in controlling the level of phenotypic variability within a population, which is directly tied to cell-fate decisions. As such, substantial efforts have been directed towards quantitatively modeling the effects of various post-transcriptional mechanisms on the strength of fluctuations in protein levels (noise). However, the corresponding effects of post-transcriptional regulation on the occurrence of rare events corresponding to large deviations are far less explored and have only been considered for a special model. Here, we take a general model of post-transcriptional regulation and apply the partitioning of Poisson arrivals (PPA) framework to map it onto a model that resembles promoter-based regulation of transcription, leading to a general framework to obtain objects of interest in large deviations (i.e. large deviation rate function for quantifying the likelihood of observing rare protein production rates and the corresponding driven process that characterizes the system dynamics conditional on the rare event) for models of post-transcriptional regulation directly from prior results for promoter-based models. The results derived create new avenues to analyze rare events in general models of post-transcriptional regulation pertaining to various different biological settings. 
\end{abstract}

\date{\today}
\maketitle

\section{Introduction}
Gene expression and its regulation are intrinsically stochastic processes that drive random fluctuations in both mRNA and protein copy numbers \cite{Raj-Cell-2008, Elowitz-Science-2002, Ozbudak-NG-2002, Golding-Cell-2005, Cai-Nature-2006}. Noise in gene expression allows for the emergence of rare phenotypes that play a dominant role in bet hedging and cell-fate decisions leading to phenotypic heterogeneity within clonal cell populations {\cite{Munsky-Science-2012, Sanchez-Science-2013}}. This occurs in many different contexts and has become a recurring theme in modern biological research. Notable examples include drug resistance in melanoma \cite{Schuh-CellSys-2020, Shaffer-Nature-2017}, sporulation in bacteria \cite{Mirouze-plos-2011}, and latent to active switching in HIV-1 infections \cite{Weinberger-Cell-2005, Kumar-PRL-2014}. Thus, there is great interest in characterizing the underlying cellular regulatory mechanisms that control rare events in stochastic gene expression at every level. 

A considerable amount of research, both theoretical and experimental, has highlighted the effects of post-transcriptional regulation in gene expression across numerous different control mechanisms such as small RNA (sRNA) binding \cite{Jia-PRL-2010, Waters-Cell-2009}, alternative splicing \cite{Modrek-NG-2002, Wang-PRE-2014}, multi-step mRNA decay \cite{Pedraza-Science-2008, Shi-BJ-2023}, nuclear export \cite{Singh-BJ-2012, Weidemann-SciAdv-2023}, etc. However, these studies mainly focus on the corresponding effects that the post-transcriptional control mechanisms have on the noise (typically quantified by some combination of the mean and variance) of the protein count; the question of how these mechanisms correspondingly control the occurrence of rare events is far less explored and understood. For example, non-coding RNAs (ncRNA) such as microRNAs and sRNAs often act as buffers of stochastic gene expression and are crucial for the canalization of a uniform phenotype in a population \cite{Levine-PlosBio-2007}. The ncRNA regulator binds to the mRNA which often leads to translational repression or coupled degradation. These binding and unbinding events with the sRNA that implement the buffering are stochastic processes themselves. Because of this, these regulatory mechanisms may sometimes fail to stabilize protein levels and if the protein in question participates in a threshold-like stress response \cite{Little-PNAS-2005} (akin to a first-order phase transition), such rare failure events will disproportionately contribute to phenotypic heterogeneity.

Given the need to develop quantitative models to characterize rare events in stochastic models of gene expression, prior work \cite{Horowitz-PhysBiol-2017} has developed an analytical framework based on nonequilibrium statistical mechanics using large deviation theory to characterize fluctuations and their probabilities conditional on the occurrence of a rare event in terms of the \textit{driven process} \cite{Chetrite-AHP-2015, Chetrite-PRL-2013, Chetrite-JSM-2015}. The driven process is a conditioning-free process that reproduces the trajectories of the original process conditioned on a rare event while retaining the same properties and represents the optimal protocol for realizing the rare event of interest (i.e. importance sampling) from a control theory standpoint \cite{Jack-EPJ-2015}. While this framework can be applied towards general models with promoter-based regulation and bursting, its applicability to models of post-transcriptional regulation is rather narrow and only indicates the net changes of protein burst sizes due to post-transcriptional regulation in the driven process \cite{Horowitz-PhysBiol-2017, Kumar-PhysBiol-2019}. For post-transcriptional control mechanisms which are all defined by multiple model parameters, it is unclear how changes in each of these \textit{individual} parameters contribute to the driven process for fluctuations in the burst size. 

In this paper, we utilize the PPA framework developed in Refs. \cite{Pendar-PRE-2013, Wong-PRE-2026} to map general models of post-transcriptional regulation (with Poisson mRNA arrivals) onto models that resemble promoter-based regulation of transcription. We then show that the PPA mapping allows us to directly generalize the prior framework for analyzing rare events \cite{Horowitz-PhysBiol-2017} to a broad class of models of post-transcriptional regulation, leading to new insights for fine-tuning rare events in gene expression models which were previously intractable.

\section{Connecting Full and Reduced Models}
\begin{figure*}
    \centering
    \includegraphics[width=\linewidth]{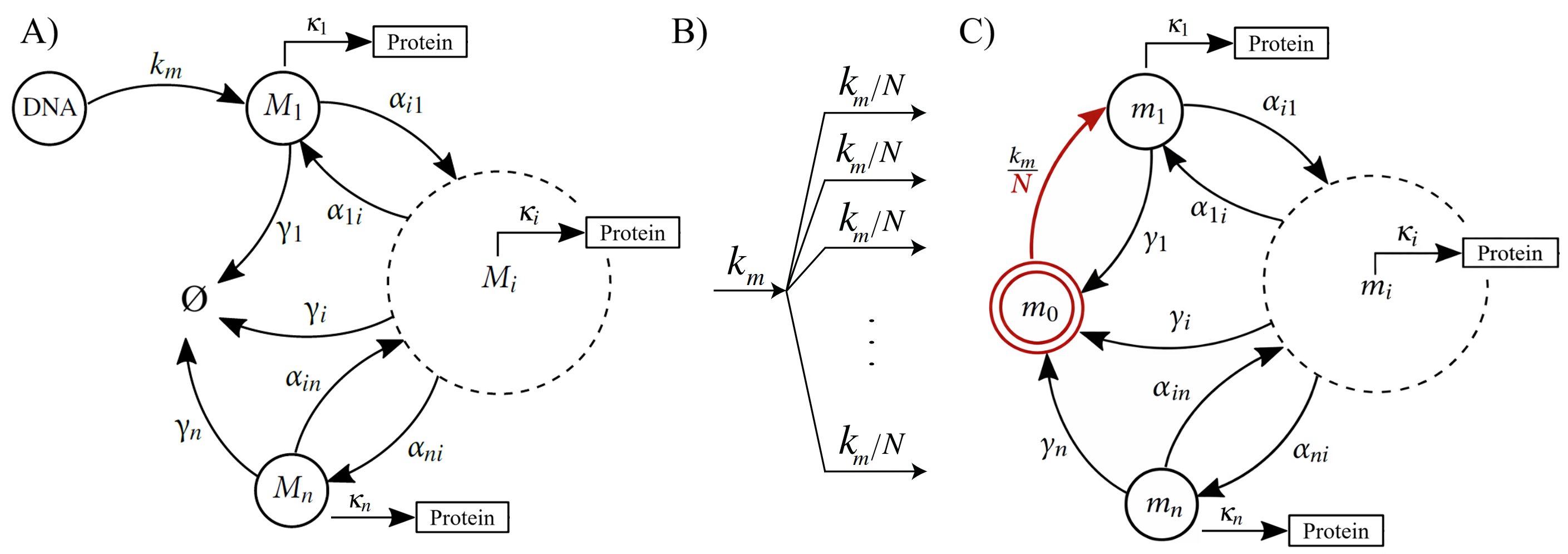}
    \caption{Adapted from our prior work \cite{Wong-PRE-2026} A) Representation of a general model of post-transcriptional regulation which we refer to as the full model. mRNA molecules may be in several mRNA states $i \in \{1, 2, ...,n\}$, with each state having distinct production and/or degradation rates: in state $i$, mRNA produce proteins at rate $\kappa_i$ and degrade at rate $\gamma_i$. We write $\alpha_{ij}$ as the transition rate from state $j$ to state $i$. B) mRNA Poisson arrivals partitioned into $N$ identical Poisson processes with a rescaled arrival rate of $k_m/N$. C) Reduced model of (A) constrained to either 0 or 1 mRNA when $N \rightarrow \infty$, resembling a promoter-based model. mRNA degradation is modeled as a transition back into the $m_0$ state.}
\end{figure*}

Consider the stochastic model of arbitrary post-transcriptional regulation depicted in Figure 1A. A single DNA molecule produces mRNA molecules at a rate of $k_m$, which are initially in the unbound state $M_1$. The mRNA can freely visit the $n-1$ other available mRNA states where in a given state $i$, the mRNA strand produces proteins at a rate of $\kappa_i$ with a corresponding degradation rate of $\gamma_i$. mRNA transitions from state $j$ to state $i$ occur at rate $\alpha_{ij}$. These states change the ability for mRNA to produce proteins and/or the rate of degradation. In such models, all production, degradation, and state-to-state transition events are assumed to occur independently of one another and taken to be counting processes. To adapt the model of post-transcriptional regulation for the analytical framework previously developed in \cite{Horowitz-PhysBiol-2017} for characterizing rare events in gene expression, we must map the system onto a model that resembles promoter-based regulation, which can be done through the PPA mapping. 

We begin by partitioning the mRNA arrivals into $N$ types (Figure 1B) with each unique type corresponding to an individual partition. Each mRNA arrival is randomly assigned to an arbitrary type $n$ drawn from $\{ 1, 2, \dots, N \}$ with uniform probability $p_n = 1/N$. From the partitioning theorem of Poisson processes, the arrival of mRNAs for each type are all independent, identical Poisson processes with a rescaled arrival rate $k_m/N$. Each partition independently contributes to the total number of mRNA arrivals $X = \sum^N_{q=1} x_q$ where $x_q$ corresponds to the random variable of the number of mRNA arrivals of the $q$-th type. This logic also applies to any random variable of interest. Thus, at any time $t$, we can map the dynamics of the full model to the dynamics of $N$ identical subsystems. 

Now, in the limit $N \rightarrow \infty$ the probability of more than 1 mRNA arrival in each subsystem can be neglected (because it is infinitely more likely that the additional arrivals get assigned to any other type). Thus, each subsystem is constrained to have at most 1 mRNA arrival with mRNA production and degradation being represented by transitions out of and into the $m_0$ state respectively. This is referred to as the \textit{reduced model} which intuitively resembles the Markovian arrival process (MAP) model of transcription (Figure 1C) \cite{Wong-PRE-2026}. Let the vector ${\mathcal{P}}(a, t) = \big\{P_{s}(a, t)\big\}_{s=0}^{n}$ specify the probabilistic state of the system for each mRNA state at time $t$ where $a$ is the random variable corresponding to the cumulative number of protein arrivals in the reduced model and $s$ indicates the underlying mRNA state. The dynamics evolves according to the system of master equations given by:
\begin{equation}
    \frac{\partial \mathcal{P}(a, t)}{\partial t} =  \bold{D}_{0} {\mathcal{P}}(a, t) + \bold{D}_{1} {\mathcal{P}}(a-1, t)
\end{equation}
where $\bold{D}_0 = \frac{k_m}{N}\boldsymbol{R}+\boldsymbol{A} - \bold{D}_1$ encodes the promoter transition dynamics with $\boldsymbol{R}_{00} = -1$, $\boldsymbol{R}_{10} = 1$, and 0 for all other elements. The block matrix $\boldsymbol{A}$ is given by 
\begin{equation}
    \boldsymbol{A} = 
    \begin{pmatrix}
        0 & \boldsymbol{\Gamma} \\ 
        0 & \boldsymbol{Q} - \text{diag} (\gamma_1, \gamma_2, ..., \gamma_n)
    \end{pmatrix}
\end{equation}
where $\boldsymbol{\Gamma} = (\gamma_1, \gamma_2, ..., \gamma_n)$ and the $n \times n$ matrix
\begin{equation}
    \boldsymbol{Q} = 
    \begin{cases}
        \alpha_{ij} \,\,\,\,\,\,\,\,\,\,\,\,\,\,\,\,\,\,\,\,\,\,\,\, i \neq j \\
        - \sum_{i \neq j} \alpha_{ij} \,\,\, i = j 
    \end{cases}.
\end{equation}
Note that the promoter transition dynamics that $\bold{D}_0$ encodes is strictly irreducible because mRNA degradation from any state is taken to be a transition back into the original $m_0$ state \cite{Wong-PRE-2026}. The elements of the $\bold{D}_1$ matrix, which encode the translation rates, are given by $\bold{D}^{ji}_{1} = \kappa_i \delta_{i,j}$. The marginal protein distribution of the reduced model is obtained from $\phi(a,t) = \bold{1} \cdot \mathcal{P}(a,t)$.

The probability distribution of the protein count in the full model $\Phi(A, t)$ can be constructed from that of each of the reduced models by 
\begin{equation}
    \Phi(A,t) = \lim_{N \rightarrow \infty} \sum_{a_1 + a_2 \ldots + a_N = A} \prod_{j=1}^N \big[ \phi_j(a_j,t) \big].
\end{equation}
Another key object of interest in the following is the moment generating function
\begin{equation}
    G(\lambda, t) = \langle e^{-\lambda A} \rangle_t = \sum_A  e^{-\lambda A} \Phi(A,t)
\end{equation}
which can be alternatively represented in terms of the reduced model by
\begin{align}
    G(\lambda,t) &= \sum_{a_1, a_2, \ldots a_N} \prod_{j=1}^N \big[  e^{-\lambda a_j} \phi_j (a_j,t)  \big] \notag \\ 
    &= \bigg( \sum_a e^{-\lambda a}  \phi(a,t)  \bigg)^N = \big[ \langle e^{-\lambda a} \rangle_t \big]^N
\end{align}
as $N \to \infty$ where the simplification comes from the fact that $a_j$ for all $j$ are identical, independently distributed random variables across all partitions, allowing us to drop their subscripts.

\section{Conditioning on Large Deviations}
The theory of large deviations seeks to quantify the probability of observing large fluctuations for a specified trajectory observable of a stochastic process in the steady state \cite{Chetrite-AHP-2015, Chetrite-PRL-2013, Jack-EPJ-2015}. In MAPs, random variables of interest may include both time-weighted (e.g. the amount of time spent in a given state of the underlying Markov chain) and counting-based observables (e.g. the number of arrival events). In the following, we will focus on the protein count observable in the reduced model unless otherwise stated. The Law of Large Numbers establishes that in the long-time limit, the \textit{rate} of the observable $a' = a/t$ will converge to its fixed average rate $\langle a' \rangle_t = \lim_{t \rightarrow \infty} a/t$ with near unit probability. However, deviations from $\langle a' \rangle_t$ in the long-time limit are possible and their probabilities can be quantified through the large deviation principle \cite{Chetrite-AHP-2015, Chetrite-PRL-2013, Jack-EPJ-2015}
\begin{equation}
    \phi(a = a't) \sim e^{-t I(a')}
\end{equation}
where $I(a')$ is the large deviation rate function. {The rate function $I(a')$ provides a measure of how unlikely it is to sustain a given protein production rate $a'$ over a long observation time where $I(a')$ vanishes at the typical production rate $\langle a' \rangle_t$ and $I(a')>0$ for atypical rates. Larger values correspond to fluctuations whose probabilities decay more rapidly with respect to the observation time. Unlike quantities such as the Fano factor (variance upon the mean) which characterize the strength of typical fluctuations about the mean, the rate function characterizes the full spectrum of steady-state production rate fluctuations and can be used to probe rare events deep within the tails of the distribution.} $I(a')$ is typically found by first constructing the scaled cumulant generating function (SCGF) $\psi(\lambda)$ for $a$
\begin{equation}
    \psi(\lambda) = -\lim_{t \rightarrow \infty} \frac{1}{t} \ln \langle e^{-\lambda a} \rangle_t,
\end{equation}
and then taking its Legendre-Fenchel transform
\begin{equation}
    I(a') = \sup_{\lambda} \{ \psi(\lambda) - \lambda a' \}
\end{equation}
which is the convex conjugate \cite{Touchette-PR-2009}. $\psi(\lambda)$ and $I(a')$ are the nonequilibrium analogs of the free energy and entropy in equilibrium systems, respectively. When $I$ is not strictly convex and admits local minima or a linear part, $\psi$ correspondingly has a nonanalyticality such as a kink. The appearance of nonanalyticalities in $\psi$ is called \textit{dynamical phase transitions} and is analogous to kinks that emerge in the free energy function characterizing phase transitions in equilibrium statistical mechanics. In recent work \cite{Inaba-PRR-2026}, we have shown that dynamical phase transitions only arise for this class of models when the promoter dynamics are reducible and represent a sharp, qualitative change in the trajectories realizing rare events. However, since the reduced models here are strictly irreducible by construction, they cannot undergo dynamical phase transitions.

Reweighting the dynamics of the MAP towards rare production events, we introduce the control parameter $\lambda$ to define the \textit{tilted} generator \cite{Chetrite-AHP-2015, Chetrite-PRL-2013, Jack-EPJ-2015} which was previously shown to be $\mathcal{D}(\lambda) = \sum_n \tilde{\bold{D}}_n(\lambda)$ \cite{Horowitz-PhysBiol-2017}. Here, $\tilde{\bold{D}}_0(\lambda)$ remains unchanged from $\bold{D}_0$ and $\tilde{\bold{D}}_1(\lambda) = \bold{D}_1e^{-\lambda}$ so $\mathcal{D}(\lambda) = \bold{D}_0+\bold{D}_1 e^{-\lambda}$. The key insight from large deviation theory is that the dominant eigenvalue of ${\mathcal{D}}(\lambda)$, which controls the long-term behavior of the system, is $-\psi(\lambda)$. To connect the dynamics of the MAP back to the full model, using Eq. (6), the SCGF of the full model is
\begin{equation}
    \Psi(\lambda) = -\lim_{t \to \infty} \frac{1}{t} \ln G(\lambda,t) = \lim_{N \to \infty} N \psi(\lambda). 
\end{equation}
Given the relations $\frac{d\psi(\lambda)}{d\lambda} = a'$ and $\frac{d\Psi(\lambda)}{d\lambda} = A' = \lim_{N \to \infty} N \frac{d\psi(\lambda)}{d\lambda}$ for a fixed value of $\lambda$, it is easy to see that $a' = A'/N$. This means conditioning the full model on a steady-state protein production rate of $A'$ is equivalent to conditioning the reduced model on $a' = A'/N$. For the rate function of the full model $\mathcal{I}(A') \sim -\lim_{t \to \infty} \frac{1}{t} \ln \Phi(A = A't)$, we similarly obtain the relation
\begin{align}
    \mathcal{I}(A') = \sup_\lambda \{ \Psi(\lambda) - \lambda A' \} = \lim_{N \to \infty} N I(a')
\end{align}
which readily follows from making use of the identities of Eq. (10) and $A'=Na'$. {The relations in Eqs. (10) and (11) exhibit an extensive scaling analogous to the thermodynamic limit. As $N\to\infty$, each reduced model partition contributes $\mathcal{O}(N^{-1})$ to the SCGF and rate function which vanish for large $N$, while the corresponding extensive quantities of the full model remain finite after combining the contribution from all $N \to \infty$ independent partitions.}

Additionally, $\mathcal{D}(\lambda)$ can also be used to characterize the driven process, which is of interest because it represents the most probable way to realize a given (possibly rare) rate $a'$ for the random variable $a$ in terms of a set of modified model parameters. The driven process reproduces the statistics of the original process \textit{conditioned} on observing $a'$ as a conditioning-free process \cite{Chetrite-AHP-2015, Chetrite-PRL-2013, Jack-EPJ-2015} and has been shown to be closest to the original process in terms of relative entropy from a control theory perspective because its dynamics retain the same structure as the original process \cite{Chetrite-JSM-2015}. Another insight from large deviation theory is that the quantities needed to characterize the driven process are encoded in the left eigenvector $\mathcal{L}(\lambda)$ of ${\mathcal{D}}(\lambda)$ corresponding to $-\psi(\lambda)$ \cite{Chetrite-AHP-2015, Chetrite-PRL-2013, Jack-EPJ-2015}. The elements of the generator of the driven process $\Delta^{ij}_n (\lambda^\ast)$ are given by the generalized Doob $h$-transform \cite{Horowitz-PhysBiol-2017}:
\begin{equation}
    \Delta^{ij}_n(\lambda^\ast) = \frac{\mathcal{L}_i(\lambda^\ast)}{\mathcal{L}_j(\lambda^\ast)}\tilde{\bold{{D}}}^{ij}_n (\lambda^\ast) + \psi(\lambda^\ast) \delta_{i,j} \delta_{n,0}
\end{equation}
with $\lambda^\ast = -\frac{dI(a')}{da'}$ being a function of the conditioned activity level $a'$. {Given that the random variable $a'$ (and by virtue $A'$ as well) is derived solely from the dynamics of protein production, the dynamics for the driven process are completely independent of the protein decay dynamics \cite{Horowitz-PhysBiol-2017} and will be the same regardless of if $\mu=0$ or $\mu$ is finite, which is why we chose to neglect it.} The driven process corresponds to another MAP with the modified set of model parameters:
\begin{align}
     \tilde{k}_m = k_m\frac{\mathcal{L}_1(\lambda^\ast)}{\mathcal{L}_0(\lambda^\ast)} ; \,\,\, \Tilde{\alpha}_{ji} = \alpha_{ji} \frac{\mathcal{L}_j(\lambda^\ast)}{\mathcal{L}_i(\lambda^\ast)}; \,\,\, \tilde{\gamma}_j = \gamma_{j} \frac{\mathcal{L}_0(\lambda^\ast)}{\mathcal{L}_j(\lambda^\ast)}; \,\,\, \Tilde{\kappa}_i = \kappa_i e^{-\lambda^\ast}
\end{align}
where $N \to \infty$ is to be enforced. Recalling that conditioning the reduced model on $a'$ is equivalent to conditioning the full model on $A'$, it follows that the parameters of the driven process between them are identical.

The results derived in this section constitute an analytical framework for characterizing rare events for a general model of post-transcriptional regulation. In the following, we demonstrate the utility of the framework on selected biologically relevant models.

\section{Applications}
\subsection{sRNA Repression}
\begin{figure}
    \centering
    \includegraphics[width=0.69\linewidth]{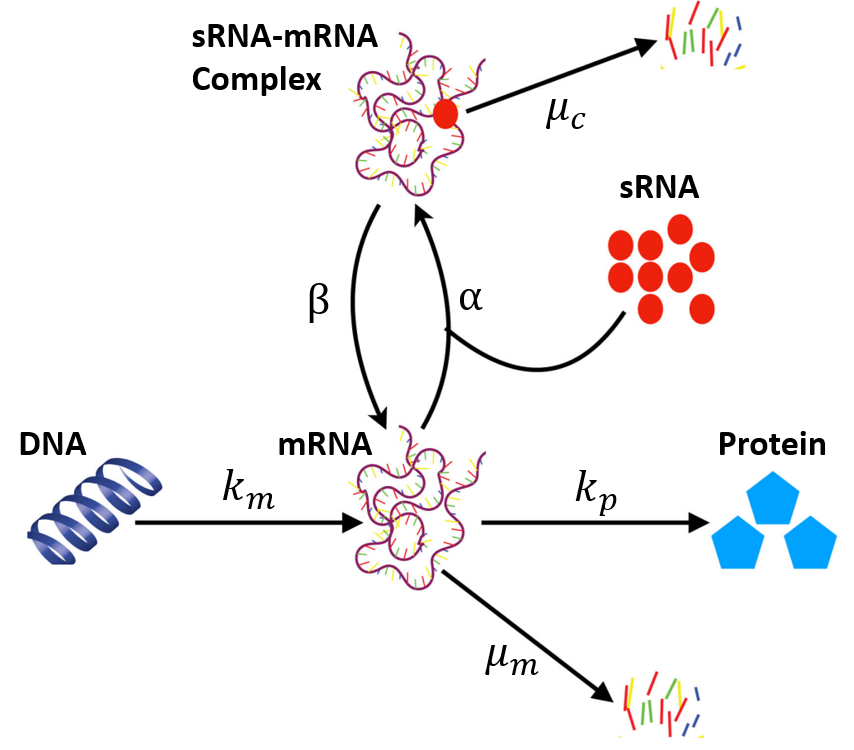}
    \caption{Adapted from Ref. \cite{Kumar-PhysBiol-2019}. Kinetic scheme for full repression of translation by a sRNA binding regulator.}
\end{figure}

Consider the model of full repression of translation by sRNA binding depicted in Figure 2. The unbound (free) mRNA are produced at a rate of $k_m$, degrade at rate $\mu_m$, and translate at rate $k_p$. A single unbound mRNA molecule binds to a sRNA molecule at rate $\alpha$, forming an mRNA-sRNA complex which fully represses translation. The mRNA-sRNA complex unbinding rate is $\beta$. We assume that individual mRNA-sRNA binding and unbinding events have a negligible effect on the overall sRNA concentration so we take $\alpha$ to be constant. The mRNA-sRNA complex degrades at rate $\mu_c$. This model was previously analytically studied in Refs. \cite{Wong-PRE-2026, Jia-PRL-2010, Kumar-PhysBiol-2019} in which it was interestingly shown that the protein burst sizes (i.e. number of proteins produced by a singular mRNA before degradation) remain geometrically distributed from the unregulated model with a modified mean and can be expressed in terms of Little's Law \cite{Elgart-PRE-2010} as
\begin{equation}
    b = k_p \langle T \rangle; \, \, \, \langle T \rangle = \frac{1}{\mu_{\text{eff}}} = \bigg[ \mu_m + \frac{\mu_c \alpha}{\mu_c + \beta} \bigg]^{-1}
\end{equation}
where $T$ is the random variable corresponding to the effective mRNA lifetime. The line of work has subsequently shown that the driven process to realize rare events in the steady-state mean protein level for both high and low activity is dominated by fluctuations in their burst sizes rather than arrival rate. However, it remains unclear whether these rare fluctuations in the burst sizes are primarily driven by changes in the translation efficiency or the effective mRNA lifetime modulated by the degree of sRNA repression. 

\begin{figure}
    \centering
    \includegraphics[width=0.8\linewidth]{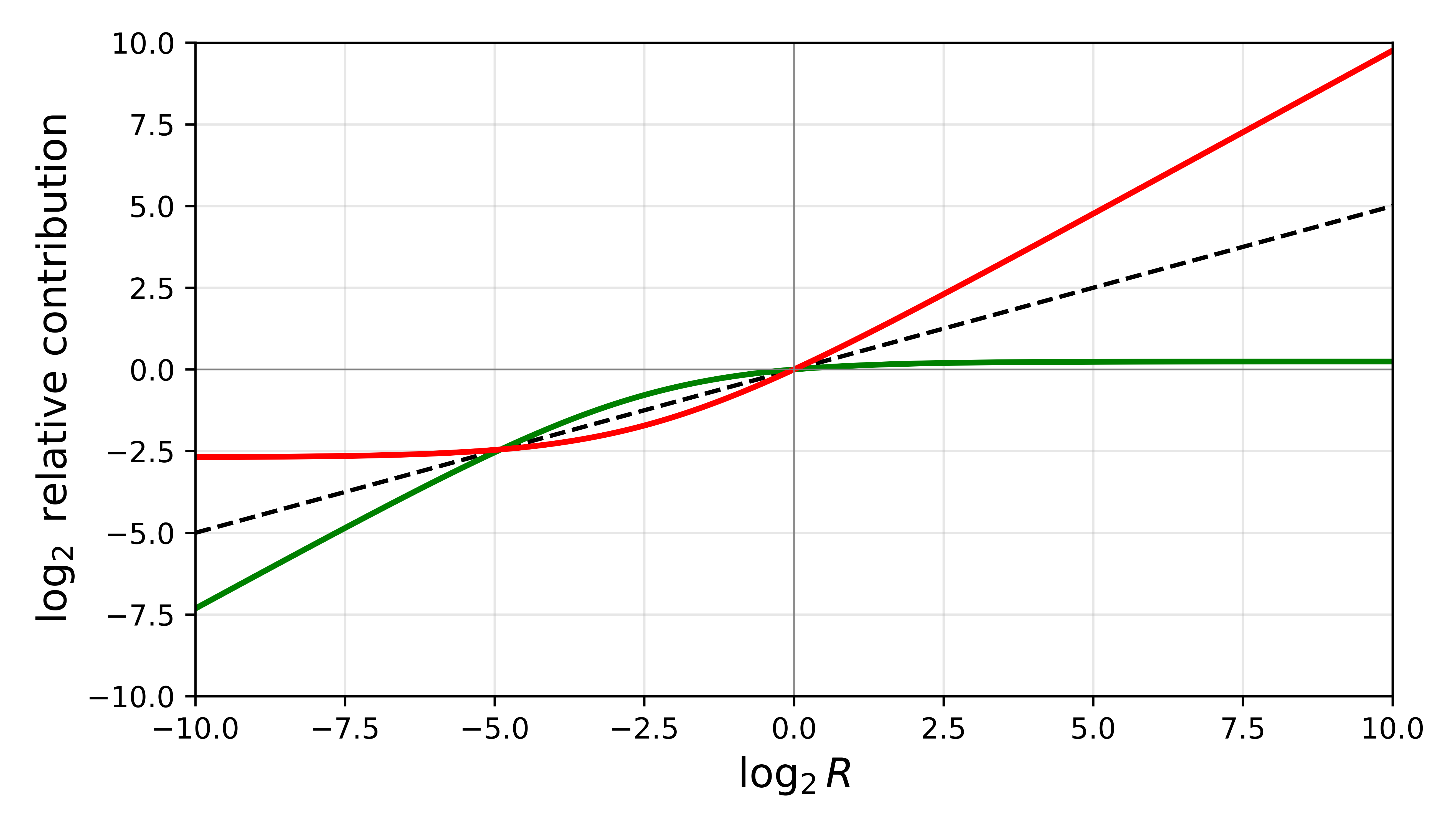}
    \caption{Relative fold change contributions to the driven process from \textcolor{ForestGreen}{$f_k$ (green)} and \textcolor{red}{$f_T$ (red)} as a function of the total fold change $R$. Dashed black line indicates equal contribution to the total fold change between $f_k$ and $f_T$. Parameters: $\alpha=\beta=\mu_m=1 \, s^{-1}; \mu_c=5 \, s^{-1}; k_m=k_p=10 \, s^{-1}$. $f_k$ and $f_T$ nontrivially intersect at approximately $(-4.895, -2.477)$ on the $\log_2$ scale.}
\end{figure}

We {define} a given fold change $R$ in the protein burst size as
\begin{equation}
    R = \bigg( \frac{\tilde{k}_p}{k_p} \bigg) \bigg( \frac{\langle \tilde{T} \rangle}{\langle T \rangle} \bigg); \,\,\, \langle \tilde{T} \rangle = \bigg[ \tilde{\mu}_m + \frac{\tilde{\mu}_c \tilde{\alpha}}{\tilde{\mu}_c + \tilde{\beta}} \bigg]^{-1}
\end{equation}
{which is the product between the fold change of the translation rate and effective mRNA lifetime.} $\tilde{k}_p$ and $\langle \tilde{T} \rangle$ are the respective quantities that characterize the driven process for realizing the fold change, which we obtain {by first diagonalizing the tilted generator for the reduced model 
\begin{equation}
    \mathcal{D}(\lambda) = \begin{pmatrix}
        -\frac{k_m}{N} & \mu_m & \mu_c \\
        \frac{k_m}{N} & -\mu_m-\alpha & \beta \\
        0 & \alpha & -k_p(1-e^{-\lambda}) - \beta - \mu_c
    \end{pmatrix}
\end{equation}
to obtain the dominant eigenvalue $-\psi(\lambda)$ and corresponding left eigenvector $\mathcal{L}(\lambda)$ and then following the formulas given in Eq. (13). We then recast the parameters of the driven process from a function of $\lambda$ to a function of the conditioned activity level $a'$ via the substitution $\lambda = -\frac{dI(a')}{da'}$. Since $a'$ also encompasses the fold change in the transcription rate unlike $R$, they are related by $a' = R\langle a'\rangle_t \frac{\tilde{k}_m}{k_m}$. To get the numerical values of $\tilde{k}_m$ and $\langle \tilde{T} \rangle$ for any desired value of $R$, we solve the preceding equation for the corresponding numerical value of $a'$ (keeping in mind that $\tilde{k}_m$ is also a function of $a'$) which we then plug into the expressions for the driven parameters.} For convenient analysis, it is helpful to recast Eq. (15) as $\log_2 R = f_k+f_T$ with $f_k = \log_2 \frac{\tilde{k}_p}{k_p}$ and $f_T = \log_2 \frac{\langle \tilde{T} \rangle}{\langle T \rangle}$. In Figure 3 we plot the driven process for $f_k$ and $f_T$ as a function of the total fold change $R$. We readily see that in the upregulated regime $R>1$, rare events in the burst size are significantly more attributable to an increase in the effective mRNA lifetime rather than translational efficiency. Interestingly, in the downregulated regime $R < 1$, the driven process for repression when $R \gtrsim 1/30$ (for our choice of model parameters) is only slightly more attributable to decreases in $f_T$ instead of $f_k$. However, for more extreme levels of downregulation, sharp decreases in the translational efficiency then dominate the driven process. We subsequently verified that the qualitative behavior of $f_T$ and $f_k$ as well as their nontrivial intersection remains consistent across various ratios of $k_p$ to ${\mu_{\text{eff}}}$ and is not specific to any choice of parameters.

\subsection{Non-Markovian mRNA Decay}
\begin{figure}
    \centering
    \includegraphics[width=0.69\linewidth]{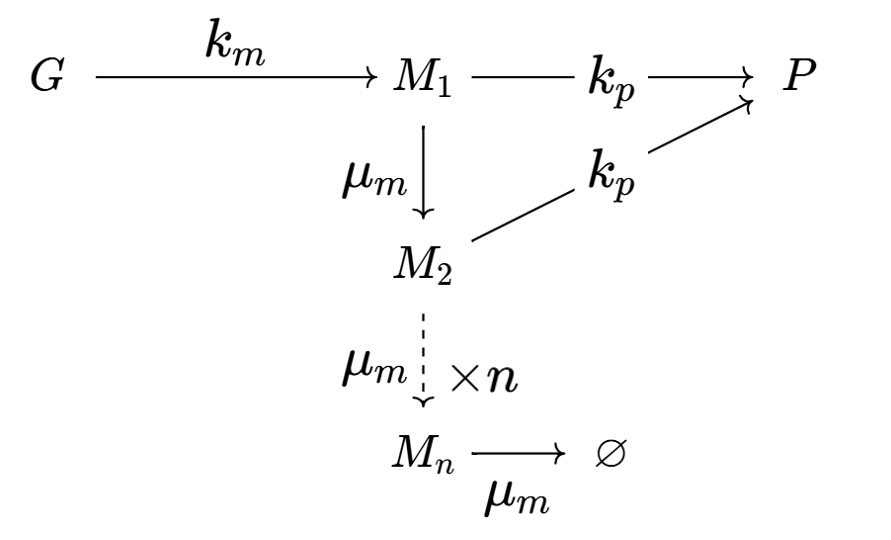}
    \caption{Adapted from our prior work \cite{Wong-PRE-2026}. Kinetic scheme for an mRNA which must undergo $n$ steps before degradation while being translationally active in intermediate steps, leading to Erlang-distributed mRNA lifetimes.}
\end{figure}

Although mRNA decay is widely modeled as a trivial, first-order process which implies exponentially distributed (memoryless) lifetimes, recent research has shown that this is not true for the majority of mRNA species {\cite{Shi-PRR-2026, Zhang-PNAS-2019, Vilk-PRR-2024}}. Thus, as depicted in Figure 4, we model mRNA decay as an arbitrary $n$-step process with the transition rate between each step being $\mu_m$ while still being translationally active at rate $k_p$, leading to mRNA lifetimes being drawn from $T \sim \text{Erlang}\big[ n, \mu_m \big]$. Setting $n=1$ recovers the Markovian case. The repetitive structure of the tilted generator for the reduced model yields a clean analytical derivation of the SCGF of the full model:
\begin{equation}
    \Psi(\lambda) = k_m\bigg[ 1- \bigg( \frac{\mu_m}{\mu_m+k_p(1-e^{-\lambda})} \bigg)^n \bigg]
\end{equation}
which we show in the Appendix. In Figure 5 we plot the corresponding rate function for $n \in \{1,2,5,10\}$ and take $\langle T \rangle$ to be five times longer than the wait-time between transcription events $\frac{1}{k_m}$, consistent with experimental observations for heavily regulated housekeeping genes in prokaryotes \cite{Tani-RNABio-2012}. For each choice of $n$ we adjusted $\mu_m$ to enforce a fixed average mRNA lifetime $\langle T \rangle = \frac{n}{\mu_m}$ to hold the mean protein level constant. From the plot, we readily see that as $n$ increases from the one-step Markovian case, the rate function monotonically narrows. This means that introducing additional mRNA degradation steps (decreasing the variability of $T$) acts as a buffer of stochastic gene expression by decreasing the probability of realizing rare protein production rates for both low and high activity levels, and vice versa. 

Notably, $\mathcal{I}(A')$ appears to monotonically narrow at a decreasing rate with respect to $n$ increasing, implying a ``diminishing marginal return" in the ability of each additional degradation step to buffer the protein level. The marginal effect of each degradation step can be quantified using the susceptibility $\chi_n(A') = \mathcal{I}_{n+1}(A') - \mathcal{I}_n(A')$, which we plot in the inset of Figure 5 for fixed values of $A'$ while keeping $\langle T  \rangle$ held constant as before and adjusting $k_p$ to achieve the corresponding value of $A'$. We see that the marginal buffering ability of a degradation step in both the upregulated and downregulated cases is substantial when there are only a few steps. However, as $n$ further increases, the marginal buffering ability rapidly monotonically decays to be negligible.

\begin{figure}
    \centering
    \includegraphics[width=0.69\linewidth]{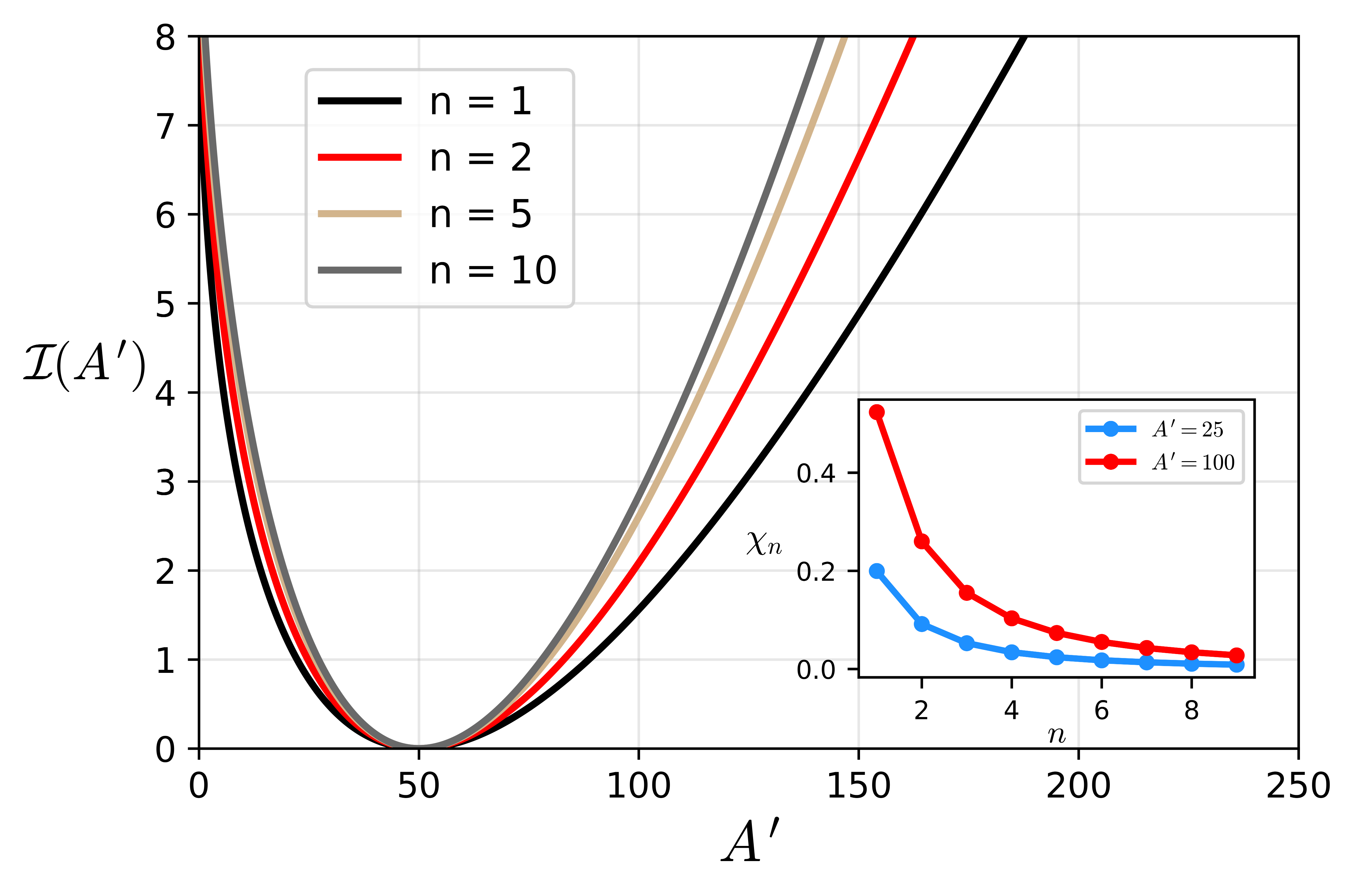}
    \caption{Large deviation rate functions plotted for $n \in \{1,2,5,10\}$ with the mean production rate held constant at $A'=50 \, s^{-1}$. Parameters: $k_m = k_p = 10 \, s^{-1}$; $\mu_m = 2n \, s^{-1}$. Inset figure: susceptibility $\chi_n(A') = \mathcal{I}_{n+1}(A') - \mathcal{I}_n(A')$ for fixed mean mRNA lifetimes and protein arrival rates $A' \in \{ 25\, s^{-1}, 100\, s^{-1} \}$. Parameters remain the same, except $k_p$ is adjusted to achieve the desired values of $A'$.}
\end{figure}

\section{Discussion and Conclusion}
In this paper, we have used the PPA mapping \cite{Pendar-PRE-2013, Wong-PRE-2026} to extend the large deviations framework for promoter-based models of gene expression \cite{Horowitz-PhysBiol-2017} to the general setting of post-transcriptional regulation. This extension provides a quantitative framework for analyzing rare events {in the protein production rate} for models of stochastic gene expression undergoing general post-transcriptional regulation by providing exact formulae for calculating rare event probabilities via the rate function and the parameters of the driven process characterizing the system dynamics conditioned on rare fluctuations. {The rate function and driven process provide complementary insights: changes in $\mathcal{I}(A')$ quantifies how the probabilities of realizing rare values of $A'$ are affected with respect to the choice of parameters (Section 4.2) while the driven process prescribes the changes in the underlying kinetic parameters to make a specified (presumably rare) value of $A'$ typical, which represents most probable mechanism through which that fluctuation is realized (Section 4.1).} 

While post-transcriptional regulation can be analyzed in the original large deviations framework for promoter-based regulation and bursting \cite{Horowitz-PhysBiol-2017, Kumar-PhysBiol-2019}, only the net change in the protein burst size (due to post-transcriptional regulation) for the driven process can be obtained. Since post-transcriptional control mechanisms are characterized by multiple parameters, the key advantage of our framework is that it allows for more detailed insights into the control of rare events by specifying the contribution of each individual parameter to the driven process, as demonstrated in Section IV. Furthermore, given that exact analytical distributions for all but the simplest gene expression models depicting post-transcriptional regulation are typically inaccessible except in special limits \cite{Pendar-PRE-2013,Kuntz-JCP-2019, Shahrezaei-PNAS-2008}, the main focus of the literature has been on analytically quantifying the noise of the protein levels via the first two moments \cite{Nossan-BPJ-2024, Jia-PRL-2011}. Although these quantities (e.g. Fano factor) specify typical fluctuations around the average expression level and are useful for giving insights into controlling the degree of phenotypic heterogeneity in a cell population, they provide limited understanding of the occurrences of rare events. This limitation is apparent for cell populations that exhibit threshold-like phenotypic responses where the phenotype does not continuously track with the protein level \cite{Little-PNAS-2005}. Our framework is particularly useful in this setting because it allows us to condition on any chosen activity level to probe the relevant rare production regimes at the threshold. By identifying both the tail probabilities of these rare events and the corresponding driven process, our framework provides a mechanistic route for designing optimal post-transcriptional control strategies to fine-tune the occurrences of rare events and thus the degree of phenotypic variability, which is of interest in synthetic biology.

A limitation of our framework is that mRNA arrivals must follow a Poisson process and we cannot incorporate promoter-based regulatory mechanisms such as transcriptional bursting of mRNA arrivals \cite{Kumar-PlosCompBio-2015, Raj-PLOSBiol-2006, Suter-Science-2011, Larsson-Nature-2019} which is a significant source of noise. {Our present formulation also does not account for stochastic partitioning of mRNA/protein molecules between daughter cells during mitosis (where mRNA/protein molecules are often unevenly split, referred to as partitioning errors); such mechanisms further propagate cellular heterogeneity \cite{Huh-PNAS-2011, Jedrak-PRE-2019, Beentjes-PRE-2020} and extending the PPA mapping into this setting is an open problem.} The formulation for both of these extensions are currently unclear. Nonetheless, the results obtained have opened new avenues for understanding rare events in stochastic gene expression modulated by post-transcriptional regulation with applications ranging from importance sampling to designing optimal cellular regulatory strategies to control the emergence of rare phenotypes.

\section*{Acknowledgments}
We thank Niraj Kumar and Rahul Kulkarni for helpful discussions and feedback in preparing the manuscript for submission.

\setcounter{equation}{0}
\renewcommand{\theequation}{A\arabic{equation}}
\section*{Appendix: Derivation of $\Psi(\lambda)$ for Erlang-distributed mRNA Lifetimes}
Consider the tilted generator of the reduced model {which takes the form of an irreducible Metzler matrix}
\begin{equation}
    \mathcal{D}(\lambda) = 
    \begin{pmatrix}
        -\frac{k_m}{N} & 0 & 0 & 0 &\dots & 0 & \mu_m \\
        \frac{k_m}{N} & -f(\lambda) & 0 & 0 &  &  & 0 \\
        0 & \mu_m & -f(\lambda) & 0 &  &  & 0 \\
        0 & 0 & \mu_m & -f(\lambda) &  &  & 0 \\
        \vdots &  &  & \ddots & \ddots &  & \vdots \\
        0 &  &  &  & \mu_m & -f(\lambda) & 0  \\
        0 & 0 & 0 & \dots & 0 & \mu_m & -f(\lambda)
    \end{pmatrix}
\end{equation}
where $f(\lambda) = \mu_m + k_p(1-e^{-\lambda})$ and denote $-\psi(\lambda)$ and $\mathcal{L}(\lambda)$ to be the dominant eigenvalue and corresponding left eigenvector respectively. Based on the identity $\mathcal{L}(\lambda)\mathcal{D}(\lambda) = -\psi(\lambda) \mathcal{L}(\lambda)$, the first, $n$-th, and $(n+1)$-th columns give the relations 
\begin{align}
    \frac{k_m}{N}\mathcal{L}_1 &= \bigg(-\psi(\lambda) + \frac{k_m}{N} \bigg) \mathcal{L}_0; \\
    \mu_m \mathcal{L}_n &= \big(-\psi(\lambda)+f(\lambda) \big) \mathcal{L}_{n-1}; \\
    \mu_m \mathcal{L}_0 &= \big(-\psi(\lambda) + f(\lambda) \big) \mathcal{L}_n.
\end{align}
Eq. (A3) is recursive and can be iterated as
\begin{equation}
    \mathcal{L}_n = \bigg( \frac{-\psi(\lambda)+f(\lambda)}{\mu_m} \bigg)^{n-1} \mathcal{L}_1.
\end{equation}
Now, substituting this expression for $\mathcal{L}_n$ in Eq. (A4), then substituting the resulting expression for $\mathcal{L}_0$ (in terms of $\mathcal{L}_1$) into Eq. (A2) and rearranging,
\begin{equation}
    \psi(\lambda) =  \frac{k_m}{N} \bigg[ 1 - \bigg( \frac{\mu_m}{-\psi(\lambda) + \mu_m + k_p(1-e^{-\lambda})} \bigg)^n \bigg].
\end{equation}
Using the identity in Eq. (10) of the main text and noting that $\psi(\lambda) \sim N^{-1}$, the SCGF of the full model is
\begin{equation}
    \Psi(\lambda) = k_m \bigg[ 1- \bigg( \frac{\mu_m}{\mu_m+k_p(1-e^{-\lambda})} \bigg)^n \bigg]
\end{equation}
where the $\psi(\lambda)$ term from the denominator has vanished after enforcing $N \to \infty$. {It remains to check that we have actually obtained the dominant eigenvalue of $\mathcal{D}(\lambda)$. By extensions of the Perron-Frobenius theorem for Metzler matrices \cite{MacCluer-SIAM-2000}, the dominant eigenvalue of $\mathcal{D}(\lambda)$ uniquely corresponds to a strictly positive left eigenvector provided that $\mathcal{D}(\lambda)$ is irreducible. Substituting for $\mathcal{L}_1$ into Eq. (A2) using Eq. (A5) and then for $\mathcal{L}_n$ using Eq. (A4) in the resulting expression, 
\begin{equation}
    \frac{k_m}{N} \mu_m^n = \bigg( -\psi(\lambda) + \frac{k_m}{N} \bigg) \big(-\psi(\lambda) + f(\lambda) \big)^n.
\end{equation} 
Recalling that $n \in \mathbb{W}$, both sides of Eq. (A8) must always be positive so it must follow that $\big(-\psi(\lambda) + f(\lambda) \big) > 0$ for $\big(-\psi(\lambda) + f(\lambda) \big)^n$ to not switch signs depending on $n$. This is needed for the equality to hold because the first factor $-\psi(\lambda) + \frac{k_m}{N}$ is fixed with respect to $n$ and thus, also constrained to be positive. Since we can always choose $\mathcal{L}_0 > 0$, it is easy to see that all other elements of $\mathcal{L}$ must share the same sign as $\mathcal{L}_0$, which implies that $-\psi(\lambda)$ is indeed the dominant eigenvalue of $\mathcal{D}(\lambda)$.}

\end{document}